\documentclass[vecphys]{svmult}		

\usepackage{makeidx}         
\usepackage{graphicx}        
\usepackage{epsfig}          
\usepackage{multicol}        
\usepackage[bottom]{footmisc}

\usepackage{amsmath}
\usepackage{amsfonts}
\usepackage{amssymb}

\makeindex             
\begin{document}

\title*{Stability of mode-locked kinks in the ac driven
and damped sine-Gordon lattice}
\titlerunning{Ac driven DSG lattice}
\author{Yaroslav Zolotaryuk }

\institute{Bogolyubov Institute for Theoretical Physics\\
National Academy of Sciences of Ukraine\\
vul. Metrologichna 14-B\\
 03680 Kiev, Ukraine\\
\texttt{email:yzolo@bitp.kiev.ua}
}
\maketitle
\abstract
{
Kink dynamics in the underdamped and strongly discrete sine-Gordon
lattice that is driven by the oscillating force is studied. The investigation
is focused mostly on the properties of the mode-locked states
in the {\it overband} case, when the driving frequency lies above
the linear band. With the help of Floquet theory it is
demonstrated that the destabilizing of the mode-locked state
happens either through the Hopf bifurcation or through the tangential
bifurcation.
It is also observed that in the overband case
the standing mode-locked kink state maintains its stability
for the bias amplitudes that are by the order of
magnitude larger than the amplitudes in the low-frequency case.
}

\vspace{0.5cm}


%

\section{Main Abbreviations}
\label{yzolo:sec1}

\begin{tabular}{ll}
$\circ$ SG: Sine-Gordon &
$\circ$ DSG: Discrete sine-Gordon \\[0.0ex]
$\circ$ FK: Frenkel-Kontorova~~~~ &
$\circ$ PN: Peierls-Nabarro \\[0.0ex]
$\circ$ JJ: Josephson junctions &
$\circ$ JJA: Josephson junction arrays \\[0.0ex]
\end{tabular}

\section{Introduction}
\label{yzolo:sec2}

The discrete sine-Gordon (DSG) equation, also known as the
Frenkel-Kontorova (FK) model,  is ubiquitous in condensed
matter physics \cite{fm96ap,bk98pr}. It has a wide range of
applications in the dislocation theory \cite{fk38pzs}, weak superconductivity \cite{wzso96pd,u98pd} and magnetism \cite{ms91adp}.
Among the intensively discussed problems for the DSG dynamics
the problem of the topological soliton (fluxon)
response to the ac (time-periodic) bias, remains to be important.
This interest is caused in particular by the number of technological applications based on the Josephson junction arrays (JJAs), which are
successively modelled by the DSG equation.
Properties of the small ac-biased  Josephson junctions have
been extensively studied both experimentally (starting from the
pioneering papers of Shapiro \cite{s63prl}) and theoretically
(with the focus on the phase-locking \cite{k81japI} and chaotic
regimes \cite{k81japII,k96rpp}). In particular,
the rf-biased Josephson junctions have been used as a voltage standard
\cite{k96rpp,volts}.

It is well-known \cite{pk84pd} that contrary to the continuous
sine-Gordon (SG) equation the DSG equation is non-integrable,
and, moreover, it does not possess moving kink solutions. The
ac-driven DSG lattice has two independent sources of non-integrability:
the external drive (bias) and the discreteness. Interplay of these
two sources has led to a number of interesting effects:
mode-locking to the frequency of the external drive \cite{mfmfs97prb}
and kink mobility \cite{bm91prb,fm99jpc} (including its
experimental detection in periodically modulated Josephson junctions
\cite{um01prb}), various regimes of the
dynamical chaos \cite{mfmfs97prb,z12pre}, biharmonically driven
discrete kink ratchet \cite{zs06pre,z12pre} to name a few. However, these
 studies have been performed mostly in the adiabatic,
 subband (the driving frequency lies in the gap of the linear
 spectrum) or resonant (the driving frequency lies in the
 linear band) cases. The high-frequency limit when the
driving frequency exceeds the linear wave frequency by several orders
of magnitude has been studied
 in Refs. \cite{g-jk92pla,kg-js92prb}. In these papers the inversion of the
 ground state that is based on the Kapitza pendulum effect
has been reported.
The intermediate {\it overband} case when the bias frequency
exceeds the linear wave frequency, but remains approximately of
the same order of magnitude, requires a special attention. This
frequency range
 is the natural bridge between the cases, studied in the papers,
 mentioned above. The
dynamics of topological solitary waves (kinks), especially their
linear stability in the intermediately high-frequency regime is the main
aim of this paper.

The paper is organized as follows. The model, the equations of motion
and the usage of the Floquet method for the linear stability
studies are described in the next section.
In the Section \ref{yzolo:sec4} we present the main properties
of the standing kinks in the  driven DSG lattice.
The discussion and the main conclusions are given in the last Section.

\section{The model and equations of motion}
\label{yzolo:sec3}

\subsection{The DSG equation}
\label{yzolo:sec3:ssec1}

The periodically driven and damped discrete sine-Gordon (DSG) equation is
introduced in a dimensionless form as follows:
\begin{equation}\label{yzolo:1}
  \ddot{\phi}_n - \kappa \Delta \phi_n  +  \sin \phi_n + \alpha
  \dot{\phi}_n + A \cos (\omega t) = 0, ~n = \overline{1, N}~.
\end{equation}
Here $\Delta \phi_n \equiv \phi_{n+1} -2 \phi_n + \phi_{n-1}$
is the discrete Laplacian and the dot represents the time
differentiation.
The physical meaning of the field variable $\phi_n$ depends on the
underlying physical system. In the dislocation theory it stands for the
particle displacement from its equilibrium position. In the JJ theory
 \cite{barone82} $\phi_n$ corresponds to the phase difference
 of the wave functions at the $n$th junction
\footnote{
The coupling constant
$\kappa=\sqrt{\Phi_0/(2\pi I_c L)}$ measures the discreteness of the array,
where $\Phi_0$ is the magnetic flux quantum, $L$ is inductance of
an elementary cell, and $I_c$ is the critical current of an individual
junction. The dimensionless dissipation parameter is then
$\alpha=\Phi_0/(2\pi I_c R)$, where $R$ is the resistance of an
individual junction, and the time is normalized to the inverse
Josephson
plasma frequency $1/\omega_0=\sqrt{C\Phi_0/(2\pi I_c)}$ with $C$ being
the junction capacitance.
}.

Only the periodic boundary conditions
\begin{equation}\label{yzolo:2}
\phi_{n+N}(t)=\phi_n(t)+2\pi Q,~{\dot \phi}_{n+N}(t)={\dot \phi}_n(t),
\end{equation}
which correspond to the circular JJAs, are to be considered. The
topological charge
$Q$ is an integer constant that stands for the net number of kinks trapped
in the lattice. Further on only we will study only
the case of one kink ($Q=1$).
The experiments with annular JJAs have been performed for
typical lengths $N \sim 8 - 30$ (see
Refs.~\cite{wzso96pd,u98pd,baufz00prl}). In the following, we consider
the case of an array (lattice) with $N=30$.

The dispersion law for the linear excitations (phonons) reads
\begin{equation}\label{yzolo:wl}
\omega_L(q)=\sqrt{1+4\kappa~ \sin^2\frac{q}{2} }~.
\end{equation}
Due to finiteness of the array, the wavenumber $q \in [0,2\pi)$ attains
only the
discrete set of values $q_m=2\pi m/N$, $m=\pm 1,\ldots  , \pm N$.

The regular kinks, mode-locked to the
frequency of the external bias correspond
to the limit cycles of Eq.~(\ref{yzolo:1}). On these orbits, the
average kink velocity is expressed as
$\langle v \rangle =k\omega/(2\pi l)$, where the winding numbers
 $k$ and $l$ are integer. Thus, the kink travels $k$ sites during the time $lT=2\pi
l/\omega$, so that, except for a shift in space, its profile is
completely reproduced after this time interval (in the pendulum
analogy, this orbit corresponds to $k$ full rotations of the
pendulum during $l$ periods of the external drive).

\subsection{Linear stability and the Floquet theory}

In this article our main focus will be on the kinks locked to the
external drive. In
order to understand better their properties, we will focus  on their
linear stability.
The fluxon periodic orbit is computed  by finding zeroes of the map
\begin{equation}\label{yzolo:map}
{\hat{\cal I}_{kl}}(T) {\bf X}= {\bf X} ,
\end{equation}
where the vector ${\bf X}$ consists of the dynamical variables
$\{\phi_n,{\dot \phi}_n\}_{n=1}^N$. The operator ${\hat {\cal I}}_{kl}$
 stands for the integration of the equations of motion (\ref{yzolo:1}) during
 the time $l T$ and afterwards the shift of the final
solution by $k$ sites forward if $k<0$ or backward if $k>0$.
The case $k=0$ corresponds to the fluxon pinned to a lattice site.

A fixed point of the map (\ref{yzolo:map}) is a mode-locked solution
$\{\phi_n^{(0)}(t),{\dot \phi}_n^{(0)}(t)\}_{n=1}^N$
which reproduces itself after the time $lT$
with the space shift by $k$ lattice sites backward or forward. Next, we
substitute the expansion
\begin{equation}
\phi_n(t)=\phi_n^{(0)}(t)+\varepsilon_n(t)~,
\end{equation}
into Eq. (\ref{yzolo:1}). For the case of {\it standing} kink ($k=0$)
after keeping only the linear terms, we obtain the following set
of linear ODEs with periodic coefficients:
\begin{equation}\label{yzolo:A3}
{\ddot \varepsilon}_n=-\alpha {\dot \varepsilon}_n + \kappa \Delta
\varepsilon_n-\cos [\phi_n^{(0)}(t)]\varepsilon_n~,~~ n=1,2,\ldots,N.
\end{equation}
The map
\begin{equation}
\left [ \begin{array}{c} \vec{\varepsilon}(lT) \\
{\dot {\vec \varepsilon}}(lT)\end{array} \right ]=
 {\hat M}(T) \left [ \begin{array}{c} \vec{\varepsilon}(0) \\
{\dot {\vec \varepsilon}}(0)\end{array}
\right ]
\end{equation}
 is constructed from the solutions of the system (\ref{yzolo:A3}). It
relates the small perturbations
${\vec \varepsilon}(t)=\{\varepsilon_n(t)\}_{n=1}^N$ at
the time moments $t=0$ and $t=lT$.
The $2N\times 2N$ Floquet (monodromy) matrix $\hat M$ contains all the
necessary information about the linear stability of the system.
If this matrix has at least one eigenvalue with
$|\Lambda_n|>1$ ($n=1,2,\ldots, 2N$), then the system is unstable. If for
all eigenvalues
$|\Lambda_n|\le 1$, the system is stable. It is well-known \cite{via89}
that these eigenvalues come in quadruples, so that if $\Lambda_n$
is an eigenvalue, then $\Lambda_n^*$, $R/\Lambda_n$ and $R/\Lambda_n^*$
(here $R=e^{-l\alpha \pi/\omega}$, see, for example,
Refs. \cite{mffzp01pre,mfmf01pre}) are also eigenvalues.
Thus, the Floquet multipliers lie either on the circle
of the radius $R$ (will be referred to as a {\it $R$-circle}
throughout the paper) or may depart from it after collisions.
The notable difference of the ac-driven case
from the dc-driven (autonomous) case is the absence of the degeneration
with respect to time shifts, which manifests itself in the absence
of the eigenvalue $\Lambda=1$ \cite{sm97no}.

Collision of the Floquet eigenvalues on the real axis
signals the tangential (saddle-node) bifurcation if it happens at
$\arg \Lambda=0$ or the period-doubling bifurcation if $\arg \Lambda=\pi$.
Eigenvalue collision away from the real axis means that the Hopf bifurcation
is taking place.

\section{Kinks in the high-frequency driven DSG equation}
\label{yzolo:sec4}

In this paper, we plan to compute the mode-locked limit
cycle that corresponds to the standing kink and to path-follow
it while a control parameter is changed until the cycle becomes
unstable or completely disappears. By monitoring
the Floquet eigenvalues $\Lambda_n$ one can obtain the information
about the underlying bifurcations and, consequently, about the
 unlocking process.

\subsection{The numerical scheme}
\label{yzolo:sec4:ssec1}

The scheme of the numerical studies can be described in the
following way. As a
starting iteration in the anti-continuum limit ($\kappa\equiv 0$) we
consider the kink state that can be described by the following
coding sequence
\begin{equation}
\{ \underbrace{00\cdots 00}_{n=1,\ldots,N_0},
\underbrace{2\pi \cdots 2\pi}_{n={N_0+1,\ldots,N}}\}~.
\end{equation}
This means that we start with the kink which is centered
between the $N_0$th and $N_0+1$th sites.
In should be noted that due to the translational invariance the
position of the kink center (defined by $N_0$) does not
influence on the kink properties.
The above choice of the coding sequence is
defined by the well-known (see \cite{bk98pr,pk84pd}) fact that
such configuration, sometimes referred as a bond-centered kink, is
stable for the DSG equation, in contrary to the site-centered kinks
which are
unstable.
Then the initially stable (in the $\kappa=0$ limit) mode-locked state
is continued numerically. For the numerical computation of the
fixed point of Eq. (\ref{yzolo:map}) we use the
 Newton-Raphson iterative method. With this method it is possible
to compute numerically
the respective mode-locked limit cycle for the given period $lT$
 with a desired computer precision.
For details one might consult Ref. \cite{fw98pr}, Chapter 6.1.
The advantage of this approach
is that not only attractors, but also repellers, can be computed.
Also, wrong conclusions which can be made due to sensitivity
to initial conditions can be avoided.
Once the fixed point of (\ref{yzolo:map}) is computed, $\phi_n^{(0)}$
is plugged into Eqs. (\ref{yzolo:A3}) and the Floquet matrix is
computed and diagonalized with the standard numerical methods.

\subsection{The existence diagram}
\label{yzolo:sec4:ssec2}

It is well known that discreteness causes kink pinning to the lattice \cite{pk84pd}. The kink in a lattice
 can be approximately described as a particle that
moves in the spatially periodic potential (the PN potential) with
the period that coincides with the lattice spacing. If the bias
 amplitude is weak enough, the kink will oscillate around the
 minima of the PN potential with the bias period. In other words,
 the kink oscillations are locked to the external drive. If the
amplitude (or, alternatively, the coupling) is increased till some critical value, the external drive becomes strong enough
to unlock the kink and to destroy the mode-locked state.
Intuitively, it is not hard to understand that on the parameter
plane $(\kappa,A)$ one can draw a curve $A_c=A_{c}(\kappa)$ that
marks the loss of stability of a stable mode-locked standing kink.
Suppose $\kappa$ is fixed and the mode-locked state is
continued while the amplitude $A$ is increased until this
state loses stability at $A=A_c$. At $A>A_c$, depending
on the type of the destabilizing bifurcation and
on the system parameters, different dynamical regimes can take place
such as diffusively moving kinks, mode-locked
moving kinks, quasiperiodic kinks, another standing mode-locked
kink states with the different shape or even the chaotic dynamics
of the whole array (when individual kinks cannot be identified).
The latter case corresponds to the {\it non-existing} area and will not be
discussed in this paper. The nature of the kink dynamics in the
non-pinned
area also strongly depends on the driving frequency $\omega$.

\begin{figure}
\center
\includegraphics[width=11cm]{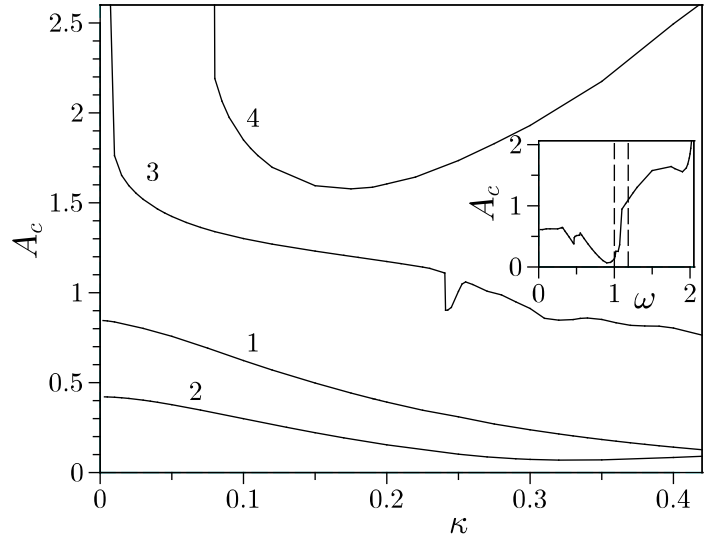}
\caption{Existence diagram of standing mode-locked kinks for
$\alpha=0.1$, $\omega=0.25$ (curve $1$), $\omega=0.7$ (curve $2$),
$\omega=1.3$ (curve $3$) and $\omega=2$ (curve $4$).
The inset shows the $A_c(\omega)$ at $\kappa=0.1$, the dashed 
vertical bars mark the edges of the linear band, $\omega_L(0)$
and $\omega_L(\pi)$.}
\label{yzolo:fig1}
\end{figure}

\begin{figure}
\center
\includegraphics[width=10cm]{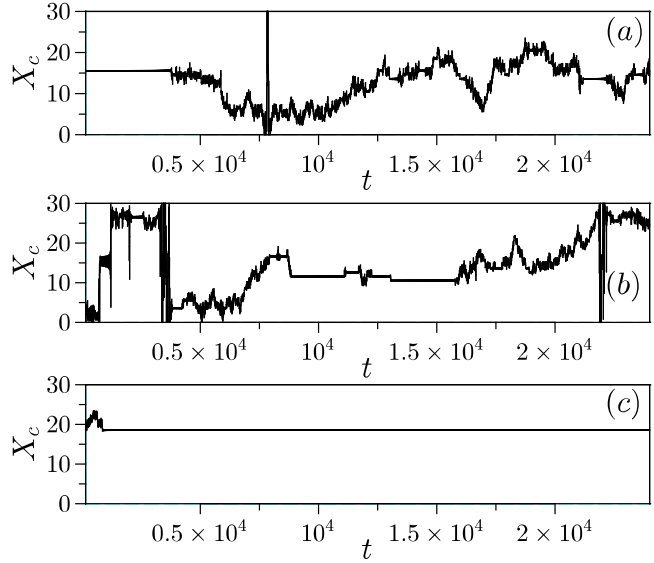}
\caption{Time evolution of the
kink center of mass $X_c=(4\pi)^{-1}\sum_{n=1}^N n(\phi_{n+1}-\phi_{n-1})$ at $\alpha=0.1$, $\omega=1.3$, $\kappa=0.1$,
and $A=1.3012$ (a), $A=1.3$ (b) and $A=1.295$ (c).}
\label{yzolo:fig1x}
\end{figure}

The existence diagrams on the $(\kappa,A)$ plane for the different values
of $\omega$ are shown in Fig. \ref{yzolo:fig1}. The
critical dependence $A_c(\kappa)$ is defined by the first bifurcation
that makes the mode-locked standing kink unstable.
In the case of {\it subband} [$0<\omega < \omega_L(q)$] frequencies the
dependence $A_c(\kappa)$ is almost monotonic (see the curves $1-2$ in
Fig. \ref{yzolo:fig1}) with the two well-defined limiting cases. In the
limit $\kappa \to \infty$, the effects of discreteness
disappear, thus $A_{c}$ should decrease. On the other hand, the decrease
of $\kappa$ means that the PN barrier becomes
stronger and thus a larger amplitude is necessary to overcome it.
As a result,  $A_{c}$  increases when $\kappa \to 0$. The exit from the
pinning area [below the curve $A_{c}(\kappa)$] can lead
to different scenarios depending on the direction of the exit.
The issue of discrete kink unlocking (depinning) in the DSG lattice
driven by the subband frequencies
has been studied in Refs.~\cite{mfmfs97prb,z12pre}.
In Fig. \ref{yzolo:fig1x}a the kink dynamics just above the $A_c$
value at $\kappa=0.1$ and $\omega=1.3>\omega_L(q)$ is shown. For these parameters
the first destabilizing bifurcation takes place at
$A_c\simeq 1.301145$, while the driving amplitude in Fig. \ref{yzolo:fig1x}a
corresponds to the slightly larger value $A=1.3012$. It
can be clearly observed that initially the kink stays pinned, but
 at $t \sim 5000$ it unlocks and begins to move chaotically.
At this point we should remark that at $A<A_c$ the staning
 mode-locked kink is not the unique solution. Typically, the
 non-uniqueness is manifested by the existence of hysteresis loops in the
 neighbourhood of $A_c$ (for the subband case see Ref. \cite{mfmfs97prb}) . This situation is demonstrated in
 Fig. \ref{yzolo:fig1x}b-c. After crossing the critical value $A_c$
 the driving amplitude is decreased back, and the chaotic moving
 solution is followed
 \footnote{It should be noted that the persistence of the diffusive
 kink solution at $A<A_c$
 has been checked for the times $t \sim 10^6$ which are much larger
 then in Fig. \ref{yzolo:fig1x}b. For the sake of clearness of the
 figure these data were not plotted.}
  to the values $A<A_c$ (Fig. \ref{yzolo:fig1x}b)
 until it falls back to the mode-locked state as shown in Fig. \ref{yzolo:fig1x}c. The hysteresis loop appears to be quite narrow,
 constituting less then $1\%$ of $A_c$.
 At $\omega=0.35$ a different scenario has been
 observed, when the stability loss at $A_c$ leads to the
 complete kink destruction and chaotic dynamics of the whole lattice.
 If this chaotic solution is followed while $A$ is decreased, it
 does not return to the mode-locked kink state. Instead, at some $A_*<A_c$
 the system falls back to the complex mode-locked state
 that includes the kink
 and one or several breathers, placed at the different lattice sites.

The dependence on the driving frequency $A_c=A_c(\omega)$ is
non-monotonic (see the inset in Fig. \ref{yzolo:fig1}). The main
resonance with the linear band can be
 clearly identified with the minimum at $\omega\simeq \omega_L(0)\equiv 1$.
In this frequency range a linear wave around the kink can be
excited at a rather small driving amplitude.
Two significantly shallow minima
at $\omega \simeq 0.5$ and $\omega \simeq 2$
appear due to the subharmonic half-frequency resonance and the
resonance with the double frequency of the linear spectrum, respectively.

The case of {\it overband} frequencies [lying above the linear band,
$\omega>\omega_L(q)$] is not  well studied yet. Let us now focus
on the curves $3-4$ in the Fig. \ref{yzolo:fig1}. It should be noted that
they are not monotonous, and, in addition they show sharp growth at
$\kappa\to 0$.
In order to understand this behaviour we study the Floquet spectrum
and the nature of the destabilizing bifurcations.

\subsection{Floquet spectrum and the destabilizing bifurcations}

In the subband (low-frequency) case the stability loss leads to the
unlocking of
the standing kink. Typically it takes place through the tangential
bifurcation \cite{mfmfs97prb} which leads to the disappearance
of the mode-locked state.  After this bifurcation the kink starts to move
in a chaotic way. More precisely, its regime belongs to the type-I intermittency. If the coupling is weak enough, the instability may lead
to the destruction of the kink state and to chaotic motion of the
whole lattice. In the neighbourhood of the main resonance $\omega \simeq 1$
 the first destabilizing bifurcation is also tangential and it takes
 place at rather small values of the amplitude.
It is caused by the resonance with the linear band and transforms
the spatially monotonic standing kink into the standing kink
with the oscillating background. Further increase of $A$ leads to the
second destabilizing bifurcation after which the kink undergoes
either depinning transition or destruction.

\begin{figure}
\center
\includegraphics[width=11.5cm]{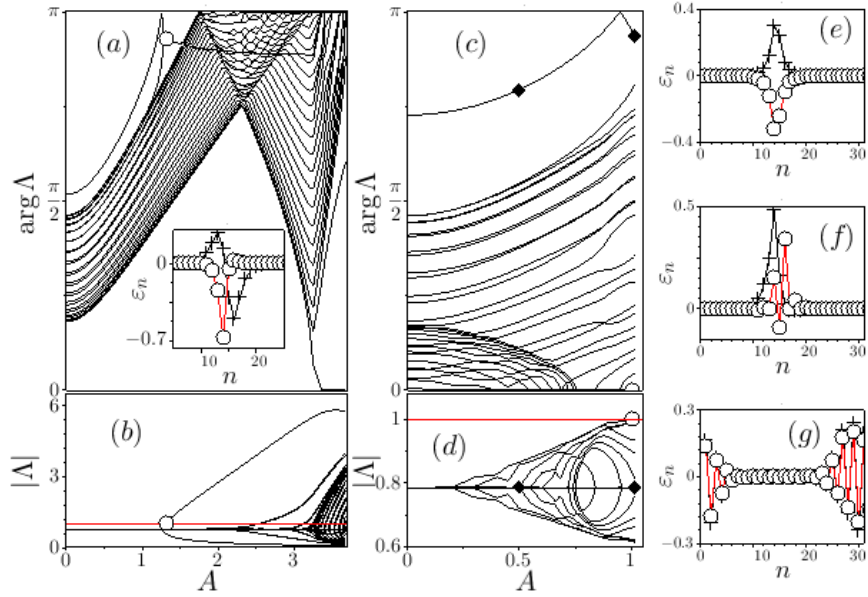}
\caption{Phases and moduli of the Floquet multipliers as a function 
of the bias amplitude for $\omega=1.3$, $\kappa=0.1$ (a-b) and 
$\kappa=0.25$ (c-d). The inset
in the panel (a) shows the unstable eigenvector at $A=1.31$.
The dashed horizontal line in the panels (b) and (d) corresponds
to the radius of the unit circle.
The unstable eigenvalues are marked by  $\circ$.
Panels (e) and (f) show the shape of the localized eigenvector
at $A=0.5$ and $A=1.015$, respectively. The respective eigenvalues
are marked by $\blacklozenge$.
Panel (g) shows the unstable eigenvector at $A=1.008$.
In panels (e)-(g) $+$ corresponds to $\mbox {Re} \epsilon_n$
and  $\circ$ corresponds to $\mbox {Im} \epsilon_n$.
The rest of the parameters is the same as in Fig. \ref{yzolo:fig1}.}
\label{yzolo:fig2}
\end{figure}

The case with $\omega=1.3$ that corresponds to the curve $3$ in the
Fig. \ref{yzolo:fig1}, can be considered either as overband or as a
resonant depending on the value of $\kappa$, which defines the upper
edge $\omega_L(\pi)=\sqrt{1+4\kappa}$ of the linear spectrum (\ref{yzolo:wl}).
Evolution of the Floquet spectrum for this case for the values of $\kappa=0.1$
and $\kappa=0.25$ is shown in Fig. \ref{yzolo:fig2}. In the first case
($\kappa=0.1$) the bias frequency lies above the linear spectrum.
In the anticontinuum limit, all the eigenvalues sit in one point
and with the growth of $\kappa$ they separate, forming two distinct
groups: the modes associated with the linear spectrum
and the internal mode(s). The linear band extends with $\kappa$
according to the dispersion law (\ref{yzolo:wl}), while
the localized eigenmode is distinctly detached from the linear band,
as can be clearly seen in Figs. \ref{yzolo:fig2}a,c.

The destabilizing bifurcation takes place at $A\simeq 1.31$ and it can be
clearly seen from the Fig. \ref{yzolo:fig2}a that this is a Hopf
 bifurcation
in which the localized mode [for the respective eigenvector shape
see the inset in the panel (a)] is involved.
Stability loss in this case causes the kink to unlock and to
start moving chaotically along the lattice.

A different bifurcation scenario is observed when $\kappa=0.25$
(Figs. \ref{yzolo:fig2}c-d). Now the resonance with the linear spectrum
occurs and the destabilizing eigenvector is delocalized as shown in the
Fig. \ref{yzolo:fig2}g. Obviously it is associated with the linear
spectrum. Indeed, one can clearly observe multiple collisions
of the eigenvalues from the linear band on the positive side
of real axis. These collisions are represented as ``bubbles'' in the
$|\Lambda_n(A)|$ dependences Fig. \ref{yzolo:fig2}d. One of these ``bubbles''
expands beyond the value $|\Lambda_n|=1$ at $A=1.008$
and causes the instability of the mode-locked state.
The further small increase
of the driving amplitude (to the value $A=1.095$)
brings the kink to the regime of chaotic diffusion.
The localized eigenvalue (Fig. \ref{yzolo:fig2}e-f)
stays distinctly detached from the linear band.

The resonance with the linear waves explains the small dip and some
oscillations in
the $A_c(\kappa)$ dependence (curve 3 in the Fig. \ref{yzolo:fig1}). In the intervals
$\kappa \in [0.241, 0.253]$ and $\kappa \in [0.33,0.335]$ the first destabilizing bifurcation
is the tangential (due to the resonance with the linear waves) and
the Hopf bifurcation occurs later,
while everywhere outside this interval it is the Hopf bifurcation
which comes first. This tangential bifurcation occurs at slightly
smaller values of $A$, therefore the respective interval on the
$\kappa$ axis is marked by the small dip.
We remind that due to the finiteness of the lattice
the linear spectrum is discrete and the external ac bias
resonates with the particular cavity modes. That is why the
tangential bifurcation takes place not for
all $\kappa \ge (\omega^2-1)/4$, but in the certain intervals of $\kappa$.
After the tangential bifurcation the kink
shape changes as it becomes surrounded by the non-decaying phonon tail.

\begin{figure}
\center
\includegraphics[width=12cm]{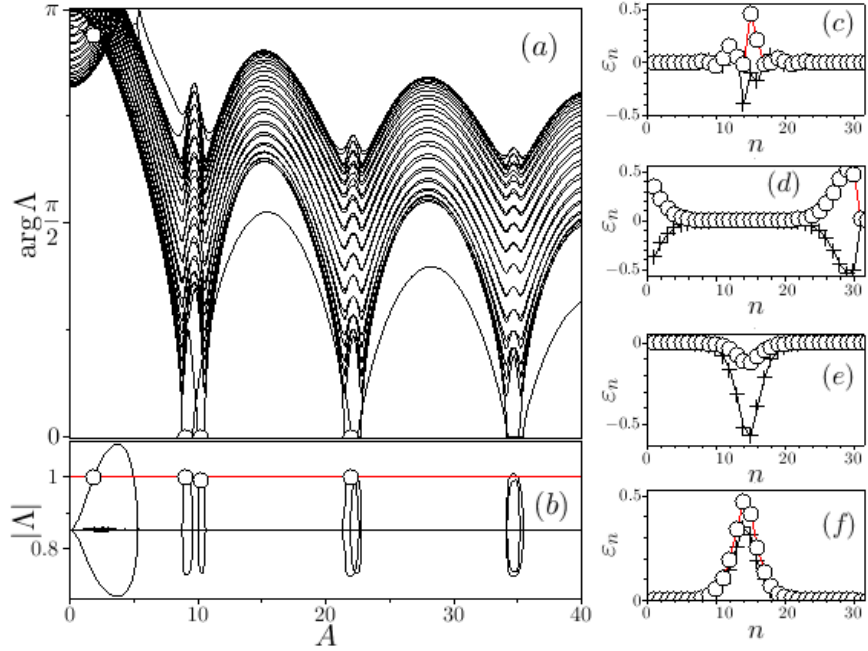}
\caption{Phases (a) and moduli (b) of the Floquet multipliers as a
function of the
bias amplitude for $\kappa=0.1$, $\omega=2$ and $\alpha=0.1$.
The dashed horizontal line in the panel (b) corresponds
to the radius of the unit circle.
The
unstable eigenvectors [$\mbox {Re} \varepsilon_n$ ($+$) and
 $\mbox {Im} \varepsilon_n$ ($\circ$)] are
shown in the panels (c) ($A=1.85$), (d) ($A=9$), (e) ($A=10.25$)
and (f) ($A=22$). The respective eigenvalues are marked by $\circ$
in the panel (a).}
\label{yzolo:fig3}
\end{figure}

\begin{figure}
\center
\includegraphics[width=8cm]{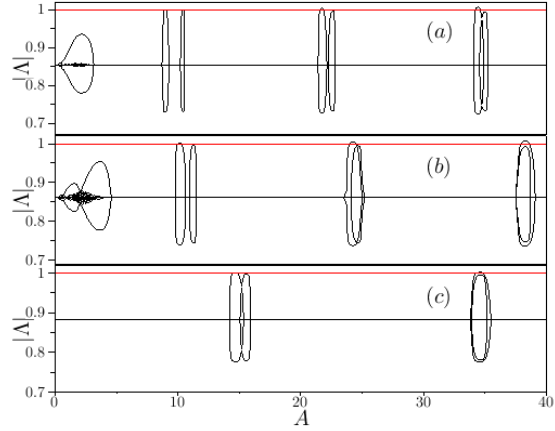}
\caption{Moduli of the Floquet multipliers as a function of the
bias amplitude for $\kappa=0.05$, $\omega=2$ (a);
$\kappa=0.1$, $\omega=2.1$ and $\kappa=0.1$, $\omega=2.5$ (c).
The dashed horizontal line marks the radius of the unit circle.}
\label{yzolo:fig4}
\end{figure}

Now we increase the bias frequency up to the value $\omega=2$. The
behaviour of the Floquet multipliers as the bias amplitude is
increased is shown in Fig. \ref{yzolo:fig3}. Similarly to the
previous case of $\omega=1.3$ the Hopf bifurcation, driven by the
spatially localized perturbation (see Fig. \ref{yzolo:fig3}c) makes
the standing kink unstable. The instability leads to the appearance of
a small localized mode on the top of the kink. The whole lattice
is no longer in the mode-locked state but in the quasiperiodic, although
the kink remains pinned.
Further increase of $A$ leads to the
second Hopf bifurcation at $A\simeq 5.25$ that restores the
mode-locked standing kink state. Inside this interval the dynamic is
quasiperiodic and no dynamical chaos has been detected.
It appears that this Hopf bifurcation can be controlled
by changing the system parameters. In particular, this can be done
by varying the coupling constant $\kappa$, the bias frequency $\omega$
or the damping constant $\alpha$. After the Hopf
bifurcation a number of tangential bifurcations take place
at the much larger values of $A$. Some of them are destabilizing
and some are not. Before discussing them we turn our attention on
the control of the first (Hopf) bifurcation.
In the Fig. \ref{yzolo:fig4} we show
how the moduli of the Floquet eigenvalues evolve with the increase of $A$
for the different parameter values.
The first case, shown in the Fig. \ref{yzolo:fig4}a corresponds to the
same frequency $\omega=2$, but the coupling is reduced to $\kappa=0.05$.
As a result, the bubble associated with the destabilizing Hopf
bifurcation is significantly reduced, and, more importantly, the collision of
the respective Floquet eigenvalues does not lead to the instability,
because these eigenvalues remain inside the unit circle. We have monitored
the intermediate cases $\kappa=0.07,0.08,0.09$ and have observed the
gradual increase of the bubble with the growth of $\kappa$ within this
interval. Now we can explain the sharp growth of the $A_c(\kappa)$
dependence when $\kappa$ is decreased (see Fig. \ref{yzolo:fig1}). It
happens because the Hopf bifurcation does not longer lead to the
kink instability, and the first destabilizing bifurcation is
the tangential bifurcation. Similar control of the destabilizing bifurcation
can be performed by increasing the damping parameter, which
decreases $R$ and removes the instability. Interestingly, the
increase of $\omega$ also reduces the Hopf bifurcation, and can completely
remove it. This is demonstrated in the Figs. \ref{yzolo:fig4}b-c, where
the $|\Lambda_n(A)|$ is plotted for $\omega=2.1$ and $\omega=2.5$. In the
case $\omega=2.1$ the ``bubble'' stays inside the unit circe, while
in the case $\omega=2.5$ it is almost unnoticeable.

\begin{figure}
\center
\includegraphics[width=9cm]{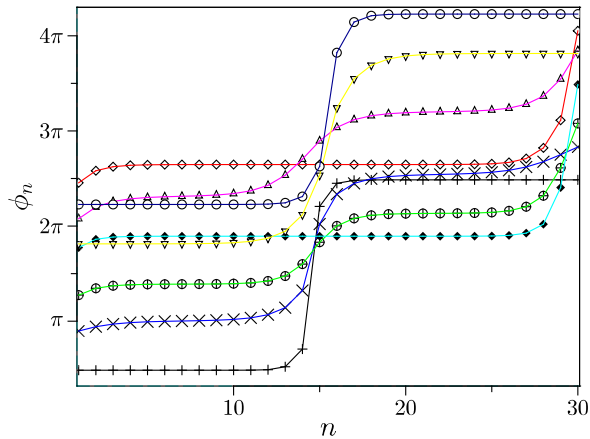}
\caption{Kink profiles for the case shown in Fig. \ref{yzolo:fig3} and for
the different ac bias amplitudes: $A=5$ ($+$), $A=9$ ($\times$),
$A=10$ ($\oplus$), $A=12$ ($\blacklozenge$), $A=20$ ($\diamond$),
$A=22$ ($\Delta$), $A=23.5$ ($\nabla$) and $A=28$ ($\circ$). Solid lines
are used as guides for an eye.}
\label{yzolo:fig5}
\end{figure}

Another interesting observation that comes from
Figs. \ref{yzolo:fig2}-\ref{yzolo:fig4} is the possibility of the
standing kink existence for the rather large bias values that exceed
the values of $A_c$ in the subband case by the order of
magnitude. Indeed, if one forgets the first Hopf bifurcation (which
can be controlled by the proper parameter choice anyway), the existence
interval of the standing kink solution stretches along the $A$-axis
interrupted only by narrow intervals where the Floquet multipliers
exit the unit circle or approach it. These intervals are represented
by the typical ``bubbles'' in the $|\Lambda_n(A)|$ dependencies
in the Fig. \ref{yzolo:fig3}b and in the Fig. \ref{yzolo:fig4}.
The amplitude of these bubbles can also be controlled by the increase
of $\alpha$ so that the instability can be removed. These underlying
bifurcations are the tangential ones and the respective unstable
perturbation in the phase space is driven by the $U$-shaped eigenvector,
as shown in Fig. \ref{yzolo:fig3}d. It means that the instability
develops in the following way: the core of the kink stays unchanged, while
the tails tend to become deformed. The second bifurcation, on the
contrary, causes the deformation in the kink core.
In the Fig. \ref{yzolo:fig5} the
change of the kink profiles with the growth of $A$ is shown. At $A=5$
the kink profile remains very much as for the usual strongly discrete
kink. Further increase of $A$ leads to the broadening of the kink
core and to the small deformation away from the core. Note that these
deformations in the tail are actually caused by the instabilities,
described in the previous paragraph. More precisely, they are associated
with the tangential bifurcation at $A\sim 9$ and the
instability direction, illustrated in
Fig. \ref{yzolo:fig3}d, prescribes exactly the same: the tail deformation.
As the amplitude increased further, the deformation
in the tails grows and the core straightens up. As a result, the kink
restores its strongly localized shape, but now it is
centered between the sites
$N-1$ and $N$ (shown by $\blacklozenge$ in Fig. \ref{yzolo:fig5}).
Also, the field variable has been increased by $2\pi$.
Further increase of the amplitude repeats the deformation scenario:
in the neighbourhood of the second set of tangential bifurcations
(around $A \sim 22 - 23.5$) the tails deform and the core
straightens up. Finally the kink re-emerges in as a strongly localized
excitation with its center placed between the sites $n=15$ and $n=16$.
Note that the initially at $A=0$ the kink position was between the
sites  $n=14$ and $n=15$. The solution has been followed up to
$A\simeq 50$ and such a transformation has been observed once again.
Thus, the increase of the bias amplitude causes a sequence
of the kink shape transformations that are driven by the tangential
 bifurcations and that shift the kink by $N/2$ sitels along the lattice.
Here we have reached the high-frequency limit, described in
Ref. \cite{g-jk92pla} where the fast-oscillating drive effectively
makes (after the averaging over the fast variables) the sine-term
in the DSG equation to
look like $J_0\left (A/[\omega\sqrt{\omega^2+\alpha^2}]\right)\sin \phi_n$
in the limit $\omega\gg 1$, and where $J_0$ is the Bessel function.
As a result, depending on the driving amplitude the stable ground may become
unstable, and back again, perfectly explaining the intermediate structures
which have flat plateau at $\phi_n \sim \pi$ [$A=10$ ($\oplus$)] or
at $\phi_n \sim 3\pi$ [$A=23.5$ ($\nabla$)]
in Fig. \ref{yzolo:fig5}.

\section{Discussion and conclusions}
\label{yzolo:sec5}

To summarize, we have studied the kink stability in the ac-driven
and damped sine-Gordon lattice. We have focused mainly on the
overband regime where the bias frequency lies above the upper edge of the
linear spectrum.

We have shown that the mode-locked high-frequency driven
standing kink is much more stable with respect to the external ac drive
comparing to the low-frequency driven kink. The critical bias
amplitude at which the mode-locked state loses its stability may
be several times larger in the high-frequency case comparing to the
low-frequency case. For example, for the coupling $\kappa=0.1$
we have $A_c\simeq 0.62$ for $\omega=0.25$ and $A_c\simeq 1.85$
for $\omega=2$. For $\kappa=0.05$ the difference is even more
drastic: $A_c \simeq 0.84$ ($\omega=0.25$) against $A_c \simeq
8.89$ ($\omega=2$). The instability of the mode-locked state
is driven in the different frequency regime
by the different bifurcations.

In the low-frequency case it is
the tangential bifurcation, associated with the internal mode,
while in the high-frequency case it is either
Hopf bifurcation, associated again with the internal mode, or the
tangential bifurcation, associated with the tail deformation. When the lattice is driven slowly, it is much
easier to drive it away from the mode-locked state because different
parts of the kink react in the different way to the perturbation: the kink
may be depinned or even destroyed if the coupling is very small.
Crossing the critical line $A_c(\kappa)$
in the low-frequency regime will almost probably lead either to the
chaotic kink diffusion or to the complete kink destruction.
The instability scenario in the high-frequency case is different. The
Hopf bifurcation transforms the mode-locked periodic kink state into
the quasiperiodic but still standing
one with the small distortion in the core. Further
increase of the driving amplitude may unlock the kink and it begins
to travel chaotically along the lattice. If one takes even higher
driving frequency and trace again the mode-locked states while
the bias amplitude is increased, such an quasiperiodic state will not turn
into the chaotic diffusive regime, but instead will be transformed
back into the mode-locked state. Thus, the Hopf bifurcation can be
controlled by the proper damping and/or frequency choice and the
instability can be arrested.

Another interesting result is the kink structural deformation,
driven by the tangential bifurcation at very large amplitudes ($A\gtrsim 10$).
Here we reach the high-frequency limit, studied in
Refs. \cite{g-jk92pla,kg-js92prb} where the ground state of the chain
alternates between the $\phi=\pi$ and $\phi=2\pi$. We would like
to note that the DSG lattice in the above-mentioned papers
has been studied at $\kappa \ge 4$ and $\omega \sim 10 - 100$
what can be considered as a high-frequency driven
 weakly discrete case, while in this article we deal with the
strongly discrete lattices at $\kappa < 1$ and in the frequency
range that exceeds the linear frequencies insignificantly: $\omega \sim 2$.

%

\printindex

\begin{thebibliography}{99.}
\expandafter\ifx\csname natexlab\endcsname\relax\def\natexlab#1{#1}\fi
\expandafter\ifx\csname bibnamefont\endcsname\relax
  \def\bibnamefont#1{#1}\fi
\expandafter\ifx\csname bibfnamefont\endcsname\relax
  \def\bibfnamefont#1{#1}\fi
\expandafter\ifx\csname citenamefont\endcsname\relax
  \def\citenamefont#1{#1}\fi
\expandafter\ifx\csname url\endcsname\relax
  \def\url#1{\texttt{#1}}\fi
\expandafter\ifx\csname urlprefix\endcsname\relax\def\urlprefix{URL }\fi
\providecommand{\bibinfo}[2]{#2}
\providecommand{\eprint}[2][]{\url{#2}}

\bibitem{fm96ap}
\bibinfo{author}{\bibfnamefont{L.~M.} \bibnamefont{Flor\'ia}} \bibnamefont{and}
  \bibinfo{author}{\bibfnamefont{J.~J.} \bibnamefont{Mazo}},
  \bibinfo{journal}{Adv. Phys.} \textbf{\bibinfo{volume}{45}},
  \bibinfo{pages}{505} (\bibinfo{year}{1996}).

\bibitem{bk98pr}
\bibinfo{author}{\bibfnamefont{O.~M.} \bibnamefont{Braun}} \bibnamefont{and}
  \bibinfo{author}{\bibfnamefont{Y.~S.} \bibnamefont{Kivshar}},
  \bibinfo{journal}{Phys. Rep.} \textbf{\bibinfo{volume}{306}},
  \bibinfo{pages}{2} (\bibinfo{year}{1998}).

\bibitem{fk38pzs}
Ya.~Frenkel and T.~Kontorova, Phys. Z. Sowietunion {\bf 13}, 1 (1938).

\bibitem{wzso96pd}
\bibinfo{author}{\bibfnamefont{S.}~\bibnamefont{Watanabe}},
  \bibinfo{author}{\bibfnamefont{H.~S.~J.} \bibnamefont{van~der Zant}},
  \bibinfo{author}{\bibfnamefont{S.~H.} \bibnamefont{Strogatz}},
  \bibnamefont{and} \bibinfo{author}{\bibfnamefont{T.~P.}
  \bibnamefont{Orlando}}, \bibinfo{journal}{Physica D}
  \textbf{\bibinfo{volume}{97}}, \bibinfo{pages}{429} (\bibinfo{year}{1996}).

\bibitem{u98pd}
\bibinfo{author}{\bibfnamefont{A.~V.} \bibnamefont{Ustinov}},
  \bibinfo{journal}{Physica D} \textbf{\bibinfo{volume}{123}},
  \bibinfo{pages}{315} (\bibinfo{year}{1998}).

\bibitem{ms91adp}
\bibinfo{author}{\bibfnamefont{H.-J.} \bibnamefont{{Mikeska}}}
  \bibnamefont{and}
  \bibinfo{author}{\bibfnamefont{M.}~\bibnamefont{{Steiner}}},
  \bibinfo{journal}{Advances in Physics} \textbf{\bibinfo{volume}{40}},
  \bibinfo{pages}{191} (\bibinfo{year}{1991}).

\bibitem{s63prl} S. Shapiro, \bibinfo{journal}{Phys. Rev. Lett.}
\textbf{\bibinfo{volume}{11}}, 80 (1963);
S. Shapiro, A. R. Janus, and S. Holly,
\bibinfo{journal}{Rev. Mod. Phys.} {\bf 36}, 223 (1964).

\bibitem{k81japI}
R.~L. Kautz, J. Appl. Phys. {\bf 52}, 3528 (1981).

\bibitem{k81japII}
R.~L. Kautz, J. Appl. Phys. {\bf 52}, 6241 (1981);
E. Ben-Jacob, Y. Braiman, R. Shainsky and
Y. Imry, Appl. Phys. Lett. {\bf 38}, 822 (1981);
R.~L. Kautz and R. Monaco, J. Appl. Phys. {\bf 57}, 875 (1985);

\bibitem{k96rpp}
R.~L. Kautz, Rep. Prog. Phys. {\bf 59} (1996) 935.

\bibitem{volts}
M.~T. Levinsen, R.~Y. Chiao, M.~J. Feldman and B.~A. Tucker,
Appl. Phys. Lett. {\bf 31}, 776 (1977);
R.~L. Kautz, Appl. Phys. Lett. {\bf 36}, 386 (1980).


\bibitem{pk84pd}
\bibinfo{author}{\bibfnamefont{M.}~\bibnamefont{Peyrard}} \bibnamefont{and}
  \bibinfo{author}{\bibfnamefont{M.~D.} \bibnamefont{Kruskal}},
  \bibinfo{journal}{Physica D} \textbf{\bibinfo{volume}{14}},
  \bibinfo{pages}{88} (\bibinfo{year}{1984}).

\bibitem{mfmfs97prb}
\bibinfo{author}{\bibfnamefont{P.~J.} \bibnamefont{Mart\'inez}},
  \bibinfo{author}{\bibfnamefont{F.}~\bibnamefont{Falo}},
  \bibinfo{author}{\bibfnamefont{J.}~\bibnamefont{Mazo}},
  \bibinfo{author}{\bibfnamefont{L.~M.} \bibnamefont{Flor\'ia}},
  \bibnamefont{and} \bibinfo{author}{\bibfnamefont{A.}~\bibnamefont{S\'anchez}},
  \bibinfo{journal}{Phys. Rev. B} \textbf{\bibinfo{volume}{56}},
  \bibinfo{pages}{87} (\bibinfo{year}{1997}).


\bibitem{bm91prb} L.~L. Bonilla and B.~A. Malomed, Phys. Rev. B {\bf 43},
11539 (1991).

\bibitem{fm99jpc}
\bibinfo{author}{\bibfnamefont{G.}~\bibnamefont{Filatrella}} \bibnamefont{and}
  \bibinfo{author}{\bibfnamefont{B.~A.} \bibnamefont{Malomed}},
  \bibinfo{journal}{J. Phys.: Condens. Matter} \textbf{\bibinfo{volume}{11}},
  \bibinfo{pages}{7103} (\bibinfo{year}{1999}).

\bibitem{um01prb} A.~V. Ustinov and B.~A. Malomed, Phys. Rev. B {\bf 64},
020302(R) (2001).

\bibitem{zs06pre}
\bibinfo{author}{\bibfnamefont{Y.}~\bibnamefont{Zolotaryuk}} \bibnamefont{and}
  \bibinfo{author}{\bibfnamefont{M.}~\bibnamefont{Salerno}},
  \bibinfo{journal}{Phys. Rev. E} \textbf{\bibinfo{volume}{73}},
  \bibinfo{pages}{066621} (\bibinfo{year}{2006}).

\bibitem{z12pre} Y.~Zolotaryuk, Phys. Rev. E {\bf 86}, 026604 (2012).

\bibitem{g-jk92pla} N.~Gr\o nbech-Jensen and Y.~S. Kivshar, Phys. Lett. A
{\bf 171} (1992) 338.


\bibitem{kg-js92prb}
Y.~S. Kivshar, N.~Gr\o nbech-Jensen, and M.~R. Samuelsen,
Phys. Rev. B {\bf 45} (1993) 7789.

\bibitem{barone82}
A.~Barone, G.~Paterno: \textit{Physics and Applications of the
Josephson Effect}, (Wiley, New York 1982).

\bibitem{baufz00prl}
\bibinfo{author}{\bibfnamefont{P.}~\bibnamefont{Binder}},
  \bibinfo{author}{\bibfnamefont{D.}~\bibnamefont{Abraimov}},
  \bibinfo{author}{\bibfnamefont{A.~V.} \bibnamefont{Ustinov}},
  \bibinfo{author}{\bibfnamefont{S.}~\bibnamefont{Flach}}, \bibnamefont{and}
  \bibinfo{author}{\bibfnamefont{Y.}~\bibnamefont{Zolotaryuk}},
  \bibinfo{journal}{Phys. Rev. Lett.} \textbf{\bibinfo{volume}{84}},
  \bibinfo{pages}{745} (\bibinfo{year}{2000}).

\bibitem{via89}
\bibinfo{author}{\bibfnamefont{V.~I.} \bibnamefont{Arnol'd}},
  \emph{\bibinfo{title}{Mathematical Methods of Classical Mechanics}}
  (\bibinfo{publisher}{Springer}, \bibinfo{address}{Berlin},
  \bibinfo{year}{1989}).

\bibitem{mffzp01pre}
\bibinfo{author}{\bibfnamefont{A.~E.} \bibnamefont{Miroshnichenko}},
  \bibinfo{author}{\bibfnamefont{S.}~\bibnamefont{Flach}},
  \bibinfo{author}{\bibfnamefont{M.~V.} \bibnamefont{Fistul}},
  \bibinfo{author}{\bibfnamefont{Y.}~\bibnamefont{Zolotaryuk}},
  \bibnamefont{and} \bibinfo{author}{\bibfnamefont{J.~B.} \bibnamefont{Page}},
  \bibinfo{journal}{Phys. Rev. E} \textbf{\bibinfo{volume}{64}},
  \bibinfo{pages}{066601} (\bibinfo{year}{2001}).

\bibitem{mfmf01pre}
\bibinfo{author}{\bibfnamefont{J.~L.} \bibnamefont{Mar\'in}},
  \bibinfo{author}{\bibfnamefont{F.}~\bibnamefont{Falo}},
  \bibinfo{author}{\bibfnamefont{P.}~\bibnamefont{Mart\'inez}},
  \bibnamefont{and} \bibinfo{author}{\bibfnamefont{L.~M.}
  \bibnamefont{Flor\'ia}}, \bibinfo{journal}{Phys. Rev. E}
  \textbf{\bibinfo{volume}{63}}, \bibinfo{pages}{066603}
  (\bibinfo{year}{2001}).

\bibitem{sm97no}
\bibinfo{author}{\bibfnamefont{J.~A.} \bibnamefont{Sepulchre}}
  \bibnamefont{and} \bibinfo{author}{\bibfnamefont{R.~S.}
  \bibnamefont{MacKay}}, \bibinfo{journal}{Nonlinearity}
  \textbf{\bibinfo{volume}{10}}, \bibinfo{pages}{679} (\bibinfo{year}{1997}).

\bibitem{bkm98pd}
\bibinfo{author}{\bibfnamefont{C.}~\bibnamefont{{Baesens}}},
  \bibinfo{author}{\bibfnamefont{S.}~\bibnamefont{{Kim}}}, \bibnamefont{and}
  \bibinfo{author}{\bibfnamefont{R.}~\bibnamefont{{Mackay}}},
  \bibinfo{journal}{Physica D Nonlinear Phenomena}
  \textbf{\bibinfo{volume}{113}}, \bibinfo{pages}{242} (\bibinfo{year}{1998}).

\bibitem{fw98pr}
\bibinfo{author}{\bibfnamefont{S.}~\bibnamefont{Flach}} \bibnamefont{and}
  \bibinfo{author}{\bibfnamefont{C.~R.} \bibnamefont{Willis}},
  \bibinfo{journal}{Phys. Rep.} \textbf{\bibinfo{volume}{295}},
  \bibinfo{pages}{182} (\bibinfo{year}{1998}).



\end{thebibliography}
\end{document}